\documentstyle[11pt,a4]{article}
\begin{document}



\newcounter{subequation}[equation]

\makeatletter
\expandafter\let\expandafter\reset@font\csname reset@font\endcsname
\newenvironment{subeqnarray}
  {\def\@eqnnum\stepcounter##1{\stepcounter{subequation}{\reset@font\rm
      (\theequation\alph{subequation})}}\eqnarray}%
  {\endeqnarray\stepcounter{equation}}
\makeatother


\newcommand{\ga}{\alpha}
\newcommand{\gb}{\beta}
\newcommand{\gc}{\gamma}
\newcommand{\gcp}{\gamma^\prime}
\newcommand{\gd}{\delta}
\newcommand{\gep}{\epsilon}
\newcommand{\gl}{\lambda}
\newcommand{\gL}{\Lambda}
\newcommand{\gk}{\kappa}
\newcommand{\go}{\omega}
\newcommand{\gp}{\phi}
\newcommand{\gs}{\sigma}
\newcommand{\gt}{\theta}
\newcommand{\gC}{\Gamma}
\newcommand{\gD}{\Delta}
\newcommand{\gO}{\Omega}
\newcommand{\gT}{\Theta}
\newcommand{\gvp}{\varphi}


\newcommand{\be}{\begin{equation}}
\newcommand{\ee}{\end{equation}}
\newcommand{\ba}{\begin{array}}
\newcommand{\ea}{\end{array}}
\newcommand{\bea}{\begin{eqnarray}}
\newcommand{\eea}{\end{eqnarray}}
\newcommand{\bes}{\begin{eqnarray*}}
\newcommand{\ees}{\end{eqnarray*}}
\newcommand{\bsea}{\begin{subeqnarray}}
\newcommand{\esea}{\end{subeqnarray}}
\newcommand{\lra}{\longrightarrow}
\newcommand{\lms}{\longmapsto}
\newcommand{\ra}{\rightarrow}
\newcommand{\pa}{\partial}


\newcommand{\N}{\mbox{I\hspace{-.4ex}N}}
\newcommand{\C}{\mbox{$\,${\sf I}\hspace{-1.2ex}{\bf C}}}
\newcommand{\Cs}{\mbox{$\,${\sf I}\hspace{-1.2ex}C}}
\newcommand{\Z}{\mbox{{\sf Z}\hspace{-1ex}{\sf Z}}}
\newcommand{\R}{\mbox{\rm I\hspace{-.4ex}R}}
\newcommand{\1}{\mbox{1\hspace{-.6ex}1}}


\makeatletter
\newcommand{\Tr}{\mathop{\operator@font Tr}\nolimits}
\newcommand{\me}{\mathop{\operator@font e}\nolimits}
\newcommand{\th}{\mathop{\operator@font th}\nolimits}
\newcommand{\ch}{\mathop{\operator@font ch}\nolimits}
\newcommand{\sh}{\mathop{\operator@font sh}\nolimits}
\makeatother

\newcommand{\la}{\wi{\Tr J^2}}
\newcommand{\DI}[1]{\mbox{$\displaystyle{#1}$}}
\newcommand{\wi}[1]{\widehat{#1}}



\thispagestyle{empty}

\hbox to \hsize{%
  \vtop{\hbox{      }\hbox{     }} \hfill
  \vtop{\hbox{DAMTP-R-98-12}}}

\vspace*{1cm}

\bigskip\bigskip\begin{center}
{\bf \Huge{Quantization of the Reissner--Nordstr\"{o}m Black Hole }}
\end{center}  \vskip 1.0truecm
\centerline{{\bf
Peter Breitenlohner\footnote{e-mail: peb@mppmu.mpg.de},
Dieter Maison\footnote{e-mail: dim@mppmu.mpg.de}}}
\vskip3mm
\centerline{Max-Planck-Institut f\"{u}r
 Physik, Werner-Heisenberg-Institut,}
\centerline{F\"ohringer Ring 6, 80805 Munich, Germany}
\vskip3mm
\centerline{{\bf
Helia Hollmann\footnote{e-mail: hollmann@damtp.cam.ac.uk}}}
\centerline{Department of Applied Mathematics and Theoretical Physics,}
\centerline{Silver Street, Cambridge CB3 9EW, England}
\vskip 2cm
\bigskip \nopagebreak \begin{abstract}
\noindent

The Reissner--Nordstr\"{o}m family of solutions can be understood
to arise from the spherically symmetric sector of a nonlinear
SO(2,1)/SO(1,1) sigma model coupled to three dimensional Euclidean
gravity. In this context a group theoretical quantization is
performed. We identify the observables of the theory and
calculate their spectra.

\end{abstract}

\newpage\setcounter{page}1

\section{Introduction}

The Einstein-Maxwell (EM) theory is the simplest member of a class of
gravity theories involving Abelian vector fields, whose dimensional
reduction to three dimensions can be dualized to a nonlinear sigma
model~\cite{BreMaiGib88}.
Depending on the signature of the Killing vector (KV) field generating the
symmetry along the reduced dimension there results either a three dimensional
Euclidean theory with a non-compact pseudo-Riemannian target space
(timelike KV)
or a Minkowskian one with a non-compact Riemannian target space (spacelike
KV).
In particular in the case of EM one obtains the coset spaces
SU(2,1)/S[U(1,1) $\times$
U(1)] resp.\ SU(2,1)/S[U(2) $\times$ U(1)]. The first case, which will be
our starting point for the quantization of the RN family, is a
4--dimensional
pseudo-Riemannian symmetric space of signature $(++--)$.
This space contains two important subspaces SU(1,1)/U(1) corresponding to
solutions of pure gravity without EM field and SO(2,1)/SO(1,1) corresponding
to static solutions.

In order to simplify matters
even more we shall restrict ourselves to spherically symmetric solutions, an
assumption justified by the spherical symmetry of the classical RN
solutions as well as by a general theorem of Israel~\cite{Isr68}
stating the spherical symmetry of any single static black hole of the EM
theory.
Under these assumptions the solutions depend only on one (arbitrary) radial
variable $\rho$
and considered as a dynamical system the field theory is replaced by a
mechanical system with a constraint. Taking this view-point the
$\rho$--dependence of the solution is obtained evolving the mechanical
system with respect to ``time'' $\rho$. The corresponding Hamiltonian
involves the Casimir operator of SO(2,1) expressed by the Noether
currents of the sigma model.
These currents are the ``observables'' of the model. The quotation marks
are meant to indicate that the concept of observables is adapted to the
dimensionally reduced theory. In order to arrive at the sigma model
structure one has to use adapted coordinates and to restrict oneself 
to coordinate
(gauge) transformations respecting this choice. Correspondingly the concept
of gauge invariance is somewhat looser compared to the general covariance of
the full theory. Requiring self-adjointness of the currents expressed as
differential operators on the configuration space
SO(2,1)/SO(1,1) naturally leads to unitary representations of SO(2,1)
on a suitable $L^2$-space over the
coset space. This provides a rather rigid group theoretical framework for
the quantization of the model already employed for the vacuum theory
in~\cite{Hol96}, \cite{HolP96}.

The Wheeler--DeWitt equation of this ``mini-superspace'' model
equation splits into a group theoretic and a 1--dimensional gravity part.
Its solution requires the spectral decomposition of the Casimir
on the $L^2$ over the coset space or its harmonic analysis to put it in
mathematical terms.
In general this is a complicated problem for non-compact groups, however, in
the case of SO(2,1)/SO(1,1) resp.  SU(2,1)/S[U(1,1)$\times$U(1)] 
the solution can be found in the mathematical literature \cite{Ros78},
~\cite{Sch87}.
For reasons of simplicity we only treat the former case
corresponding to static solutions.

Besides the Casimir we also analyze the spectra of the mass and the charge
operator, the physical observables of the model. Since the corresponding
currents do not commute, only one of them can be diagonalized simultaneously
with the Casimir. Classically the Casmir equals the square of the
``irreducible mass'' $m^2-q^2$.
In contrast to the vacuum gravity case the Casimir has also a discrete
spectrum for values with $m^2-q^2< 1/4$.

\section{Quantization of the Reissner--Nordstr\"{o}m Solutions}

As is well known
the stationary solutions of the EM theory
can be derived from the dimensionally reduced  3-dimensional action
describing
a nonlinear sigma model with target space SU(2,1)/S[U(1,1)$\times$U(1)]
coupled minimally to 3-dimensional Euclidean
gravity~\cite{Kraetal80,Mai84,BreMaiGib88}

\bea
  \label{lag}
  {\cal L} &=& \sqrt{h} \left(
    \frac{1}{2} {}^{(3)}R
      - \frac{1}{2 \gD^2} \left[ (\pa_m \gD)^2 + \go_m^2 \right]
      + \frac{1}{\gD} \left[ ( \pa_m \gvp)^2 + (\pa_m \xi)^2 \right]
                     \right) \nonumber \\
  &=& \left(
    \frac{1}{2} {}^{(3)}R ~-~ \frac{h^{mn}}{8} \Tr
    \left(
      \chi^{-1} \pa_m \chi \: \chi^{-1} \pa_n \chi
    \right)
  \right),
\eea
with $\go_m = \pa_m \go + 2 \: \xi \pa_m \gvp - 2 \: \gvp \pa_m \xi.$
$\gD$ and $\go$ are the gravitational potential $g_{tt}$ (in adapted
coordinates) and the ``twist'' potential, resp. $\xi$ is the magnetic
and $\gvp$ is the electric potential. ${}^{(3)}R$ denotes the
3--dimensional Euclidean scalar curvature and $\sqrt{h}$ is the square
root of the 3--dimensional metric.  The sigma model matrix
$\chi$ will be specified for the restriction to the Reissner--Nordstr\"{o}m
family of solutions below.
In addition to the stationarity of the solution we require its
spherical symmetry.
Then the 3--dimensional space splits into a (warped) product of
2--spheres and the positive real line. The metric can be transformed into
the standard form
\[
 ds^2=h_{mn}dx^m\;dx^n=N^2(\rho) \: d\rho^2 + r^2(\rho) \: d\Omega^2,
\]
where $\rho$ is an arbitrary radial coordinate and $d\Omega^2$ the invariant
metric of the 2-sphere.
Then the Lagrangian (\ref{lag}) simplifies to
\be
  \label{ld}
  {\cal L} = N
  \left[
    \frac{r^{\prime 2}}{N^2} + 1
    - \frac{r^2}{8N^2} \Tr
    \left(
      \chi^{-1} \chi^\prime \: \chi^{-1} \chi^\prime
    \right)
  \right].
\ee
All the fields only depend on the remaining spacelike coordinate $\rho$,
where the prime denotes the derivative with respect to $\rho$.

By reduction of the four dimensional sigma model target space to a two
dimensional subspace the Einstein-Maxwell theory can be simplified
\cite{Kraetal80} to a space of constant
curvature. In this context two of them are especially interesting:
setting $\gvp$ and $\xi$ equal to zero reduces the theory to the
stationary spherically symmetric pure gravitational case, i.e. a
SU(1,1)/U(1) nonlinear sigma model and
restricting to $\go = \xi = 0$ leads to the static truncation
of the SU(2,1)/S[U(1,1) $\times$ U(1)] nonlinear coset sigma model to
SO(2,1)/SO(1,1), the Reissner--Nordstr\"{o}m family. \cite{Hol96},
\cite{HolP96} already dealt with the quantization of the former case.
Here we discuss the quantization of the  Reissner--Nordstr\"{o}m family.

The matrix $\chi$ simplifies to
\[
  \chi =
  \left(
    \ba{ccc}
      \gD - 2 \: \gvp^2 + \DI{\frac{\gvp^4}{\gD}}
        & i \sqrt{2} \: (\gvp - \DI{\frac{\gvp^3}{\gD}})
        & -i \: \DI{\frac{\gvp^2}{\gD}} \\[0.2cm]
      i \sqrt{2} \: (\gvp - \DI{\frac{\gvp^3}{\gD}})
        & 1 - 2 \: \DI{\frac{\gvp^2}{\gD}}
        & - \sqrt{2} \: \DI{\frac{\gvp}{\gD}} \\[0.2cm]
      i \: \DI{\frac{\gvp^2}{\gD}}  & \sqrt{2} \: \DI{\frac{\gvp}{\gD}}
        & \DI{\frac{1}{\gD}}
    \ea
  \right).
\]
The Lagrangian $\cal{L}$ is invariant under SO(2,1) transformations
acting on $\chi$.
The matrix of Noether currents has the structure\footnote{
the scalar product is taken  with respect to the
matrix
\[
  \eta = \left( \ba{ccc} 0 & 0 & i \\
                               0 & 1 & 0 \\
                               -i & 0 & 0 \ea \right)
\]
}
\be
  \label{cur}
  J = \frac{f^2}{2N^2} \chi^{-1} \chi^\prime =
  \left(
    \ba{ccc}
      J_S & i J_H  & 0 \\
      i J_G & 0 & - J_H \\
      0 & J_G & - J_S
    \ea
  \right).
\ee
The currents are the essential  dynamical objects of the model obeying the
(Euclidean) field equations. It is not difficult to confirm that the
currents indeed form an so(2,1) algebra.

The meaning of the various elements in the Lie algebra is found
in \cite{Mai84}: $J_G$ corresponds to a gauge transformations,
which does not change the classical solution.
$J_H$ denotes a Harrison transformation, which leads to a change
of $\gD$ into $\gvp$ and $J_S$ is a scale transformation.

Imposing proper boundary conditions at infinity on the sigma model
fields there exists an asymptotic multipole expansion of $\chi$.
\[
  \chi \sim \sum_{n=0}^{\infty} \rho^{-n} ~ \chi_n(\gt, \gp).
\]
A suitable choice of coordinates leads to
the asymptotic behaviour of the fields
\[
  \gD = 1 - \DI{\frac{2m}{\rho}} + O(\DI{\frac{1}{\rho^2}}) \qquad
  \qquad \gvp = \DI{\frac{q}{\rho}} + O(\DI{\frac{1}{\rho^2}}),
\]
which is used to calculate the Noether charges of the theory:
\[
  Q = \chi_0^{-1} \chi_1 =
  \left(
    \ba{ccc}
      -2\:m & i \sqrt{2} \: q & 0 \\
      i \sqrt{2} \: q & 0 & - \sqrt{2} \: q \\
      0 & \sqrt{2} \: q & 2 \: m
    \ea
  \right).
\]
The mass $m$ corresponds to $J_S$ and the electric charge is related with
a linear combination of a Harrison transformation $J_H$ and a gauge
transformation  $J_G$.

We now switch to a modified Hamiltonian formalism. The slicing is performed
according to $\rho$, the spacelike coordinate \cite{Mar94}, \cite{Cav96},
\cite{CavdeAFil951}.
The application of the usual quantization algorithm -- associate
multiplication operators with the fields and differentiation operators
with the momenta -- requires some care. We aim at identifying
a complete set of observables. There is a one-to-one correspondence
between the observables of the theory and the initial data $m$ and $q$.
(The equations of motion define a geodesic motion on the coset  space.
The tangent vector at $\rho = \infty$ is determined by the mass and
the electric charge). Therefore we expect the currents to be the
observables of the quantized theory.
As a manifold the coset space SO(2,1)/SO(1,1) is the De--Sitter space, which
can be illustrated by a hyperboloid. Our coordinates, which are given
by the dimensional reduction cover only half of the hyperboloid
as illustrated in \cite{HawEll73} (p. 125). Therefore globally $\gD$ and
$\gvp$ are not a good choice of coordinates. We need coordinates which
cover the whole coset space. As a part of the quantization procedure we
therefore postulate a change of coordinates and in particular we consider
the coordinates $x, y, z$ defined by $x^2 + y^2 -z^2 = 1$ or coordinates
$t$ and $\chi$ with $ x = \ch t \: \sin\chi, y = \ch t \: \cos\chi$
and  $ z = \sh t$.
Let us first deal with the $t, \chi$--coordinate system.
The application of the usual quantization procedure supplemented by
the change of coordinates yields the Hamiltonian operator
\[
  \wi{H} = \frac{1}{4} \pa_f^2 + 1 - \frac{1}{8 f^2} \la, 
\]
which defines the Wheeler--DeWitt equation $\wi{H} \psi =  0$.
The current and the Casimir operators are calculated to be
\bea
\wi{J}_S &=& i \left( -\cos \chi \: \pa_t + \sin \chi \: \th t
  \: \pa_\chi \right) \nonumber \\
\wi{J}_H &=& - \frac{i}{\sqrt{2}}
  \left[ \sin \chi \: \pa_t + \left( \cos \chi \: \th t  + 1 \right)
  \: \pa_\chi \right] \nonumber \\
\wi{J}_G &=& \frac{i}{\sqrt{2}}
  \left[ \sin \chi \: \pa_t + \left( \cos \chi \: \th t  - 1 \right) \:
  \pa_\chi \right] \nonumber \\
  \la &=& - \pa_t^2 - \th t \:  \pa_t + \frac{1}{\ch^2 t } \:
\pa_\chi^2.
  \nonumber
\eea
$\wi{H}$ commutes with the currents and $\la$
\be
  \label{comrel1}
  [\wi{H}, \wi{J}_S] = 0, \qquad [\wi{H}, \wi{J}_G] = 0,
  \qquad [\wi{H}, \wi{J}_H] = 0, \qquad
  [\hat{H}, \la] = 0.
\ee
The Hamiltonian generates the gauge transformations such that
(\ref{comrel1}) defines the current operators and the Casimir operator
to be gauge invariant quantities, i.e. they are observables.
The Casimir commutes with the currents. Therefore the Casimir operator
and any linear combination of the currents can be diagonalized
simultaneously. The currents constitute an  so(2,1) algebra
\[
  [\wi{J}_S, \wi{J}_H] = \wi{J}_H, \qquad
  [\wi{J}_S, \wi{J}_G] = - \wi{J}_G, \qquad
  [\wi{J}_H, \wi{J}_G] = - \wi{J}_S.
\]
Hence it is not possible to ``measure'' the mass and the charge
of the black hole simultaneously.
The current and the Casimir operators are per construction self-adjoint
operators with respect to the SO(2,1) invariant measure. The scalar
product is given by
\[
  (\psi, \psi) = \int_0^\pi d\chi \int_{- \infty}^\infty dt
     \: \ch t \: |\psi(t, \chi)|^2.
\]
The spectrum of the Casimir operator can be read off in the paper by
Rossmann \cite{Ros78}. It consists of a discrete part, namely the
eigenvalues $\gl = \frac{1}{4} - \nu^2, \: \nu \in \frac{1}{2} N_0$
(here we assume that the wave function is single valued on the hyperboloid)
and a continuous part with eigenvalues $\gl \in (\frac{1}{4}, \infty)$.
However, it is instructive to look at the Casimir more closely. We
notice that $\hat{J}_H + \hat{J}_G$ is proportional to the
differential operator $\pa_\chi$. With the ansatz
$\psi(t, \chi) = e^{iq\chi} \: \Phi(t)$ and the transformation
$\Phi(t) \longrightarrow \frac{1}{\sqrt{\ch t}} \: \phi(t)$ the
eigenvalue
equation
\[
  (- \pa_t^2 - \th t \: \pa_t + \frac{1}{\ch^2 t } \pa_\chi^2)
    \: \psi(t, \chi) = \gl \psi(t, \chi)
\]
becomes a standard Schr\"{o}dinger operator
\[
  - \pa_t^2 \: \phi + V(t) \: \phi = (\gl - \frac{1}{4} ) \: \phi,
   \quad \mbox{with} \quad V(t) = \frac{\frac{1}{4} - q^2}{\ch^2 t}.
\]
For $|q| < 1/2$ the potential is repulsive. For $|q| > 1/2$ one
finds a potential valley and expects to get a discrete
spectrum as well. The discrete spectrum can be generated
algebraically. For this the coordinates $x, y, z$ are the best choice.
Let us define $\wi{J}_x, \wi{J}_y, \wi{J}_z, \wi{J}_+$ and $\wi{J}_-$  by
\bea
  \wi{J}_z &=& i \left( x \pa_y - y \pa_x \right), \nonumber \\
  \wi{J}_+ &=&  \wi{J}_x + i \wi{J}_y, \nonumber \\
  \wi{J}_- &=&  \wi{J}_x - i \wi{J}_y, \nonumber
\eea
with $\wi{J}_x = i (y \pa_z + z \pa_y)$ and $\wi{J}_y = i (z \pa_x + x \pa_z)$.
$\wi{J}_x, \wi{J}_y$ and $\wi{J}_z$ are the current operators adopted
to the structure of the G/H coset space. $\wi{J}_x$ parametrizes the
non--compact part of the group and is proportional to $\wi{J}_S$,
$\wi{J}_z$ belongs to the maximal compact subgroup and is the electric
charge operator. $\wi{J}_y$ corresponds to H.
In terms of the ``physical'' operators we find
\bea
  \wi{J}_z &=& \frac{1}{\sqrt{2}}
    \left( \wi{J}_G + \wi{J}_H \right), \nonumber \\
  \wi{J}_+ &=& - \wi{J}_S + \frac{i}{\sqrt{2}}
    \left( \wi{J}_G - \wi{J}_H \right), \nonumber \\
  \wi{J}_- &=& - \wi{J}_S - \frac{i}{\sqrt{2}}
    \left( \wi{J}_G - \wi{J}_H \right), \nonumber
\eea
which constitute the algebra
\[
  [\wi{J}_z, \wi{J}_+] = \wi{J}_+, \qquad \qquad
  [\wi{J}_z, \wi{J}_-] = - \wi{J}_-, \qquad \qquad
  [\wi{J}_+, \wi{J}_-] = - 2 \wi{J}_z.
\]
The Casimir operator is
\bea
  \la &=& - \wi{J}_z^2 + \frac{1}{2} \left(
    \wi{J}_+ \wi{J}_- + \wi{J}_- \wi{J}_+  \right) \nonumber \\
   &=& - \wi{J}_z^2 + \wi{J}_z + \wi{J}_+ \wi{J}_-
   \: = \: - \wi{J}_z^2 - \wi{J}_z + \wi{J}_- \wi{J}_+. \nonumber
\eea
As $\wi{J}_{\pm}^\dagger = \wi{J}_\mp$ we have
\[
  < \wi{J}_\pm \psi \:|\: \wi{J}_\pm \psi >
    \: = \: < \psi \:|\: \wi{J}_\mp \wi{J}_\pm \psi >  \quad \ge \quad 0
\]
and consequently
\[
  \gl + q^2 + q \ge 0 \qquad \mbox{and} \qquad \gl + q^2 - q \ge 0.
\]
These inequalities lead to $\gl \le \frac{1}{4}$ and to $q \ge s$ or 
$q \le - s$, where $\gl$ is defined to be equal to $s (1-s)$.
$\wi{J}_+ \psi$ is an eigenfunction of $\wi{J}_z$ but with q--value
increased by unity. In analogy to the algebra of the angular momentum
operators in quantum mechanics we find for every eigenvalue $\gl$ of 
the Casimir
operator a sequence of eigenvalues of the current operators $\wi{J}_z$.
The $q$--values are not confined in an interval but range from a 
finite value $q=s$ or $q=-s$ in integral steps to $\infty$ or $-\infty$ 
resp. 
The wave functions can be constructed explicitely. Let us label the
wavefunction by $s, q$.  With $\wi{J}_- \psi_{ss} = 0$ we find
$\psi_{ss}(t, \chi) = C \: \me^{-i s \chi} \: \ch^{-s} t$. A wave
function with given value of $s$ and $q > s$ can be constructed by successive
application of the raising operator $\wi{J}_+$.
Application of the raising operator $J_+$ on the wave functions 
$\psi_{s(1-s)}$ gives zero. Operating with $\wi{J}_-$ on the solution of 
this differential equation one derives the wave functions with given 
values of $s$ and $q < -s$. 

Classically the Reissner--Nordstr\"{o}m family of solutions can be
parametrized by the value of the Casimir operator $\la$, which is
evaluated on the the solutions equal to $m^2 - q^2$.
Due to the indefiniteness of the metric of the coset space $m^2 - q^2$
can have either sign. $m^2 -q^2 > 0$ corresponds to the
Reissner--Nordstr\"{o}m solutions, $m^2 = q^2$ gives the extrem
Reissner--Nordstr\"{o}m solutions and $m^2 - q^2 < 0$ label the so called
over extreme solutions. They are obtained by analytic continuation
from the solutions with $m^2 - q^2 < 0$ and are naked singularities.
In the quantized version model the Reissner--Nordstr\"{o}m solutions
belong to the continuous and the over extreme solutions belong
to the discrete spectrum of the Casimir operator. The extremal
Reissner--Nordstr\"{o}m solution is characterized by $\gl = \frac{1}{4}$.
In this case the potential of the standard Schr\"{o}dinger problem
vanishes identically.

To summarize: dimensional reduction of the Einstein--Maxwell theory
with respect to a timelike KV yields a 4--dimensional
Pseudo--Riemannian symmetric space. The subspace of static solutions
is Pseudo--Riemannian as well. In contrast to the pure gravity
coset space, which is Riemannian, the spectrum of the Casimir operator
consists of a continuous and a discrete part. The usual quantization
procedure has to be supplemented by a change of coordinates.
As the mass operator refers to a non--compact direction in the
Lie algebra the spectrum is continuous.
The charge operator belongs to a compact direction and has a discrete
spectrum imposing the wave function to be single valued on the
hyperboloid. But as the group SO(2,1) is not simply connected we do 
not know up to now of any method to decide whether the spectrum 
has integral or fractional
values. Allowing for the limiting case of the infinite covering group
the spectrum would even be continuous.
We should like to mention that the papers by Rossmann and Schlichtkrull
\cite{Ros78}, \cite{Sch87} provide the mathematical tools to quantize 
the complete stationary sector of the Einstein--Maxwell theory and to deal 
with higher dimensional hyperbolic spaces.


\end{document}